\documentclass[prl,reprint,showpacs,amsmath,amssymb, superscriptaddress]{revtex4-1}
\usepackage{graphicx}
\usepackage[utf8]{inputenc}
\usepackage{amsmath}
\usepackage{amssymb}
\usepackage[mathscr]{eucal}
\usepackage{hyperref}
\usepackage{color}

\newcommand{\Jpr}{Phys. Rev.}
\newcommand{\Jprl}{Phys. Rev. Lett.}
\newcommand{\Jpra}{Phys. Rev. A}
\newcommand{\Jprb}{Phys. Rev. B}

\newcommand{\Jnature}{Nature (London)}
\newcommand{\Jnatphys}{Nature Phys.}

\newcommand{\Jnjp}{New J. Phys.}
\newcommand{\Jepb}{Eur. Phys. J. B}
\newcommand{\EPJD}{Eur. Phys. J. D}
\newcommand{\EPL}{Europhys. Lett.}

\newcommand{\Jmp}{J. Math. Phys.}
\newcommand{\JTAP}{IEEE Trans. Antennas Propag.}
\newcommand{\JRQE}{Radiophys. Quantum Electron.}
\newcommand{\JOSAA}{JOSA A}
\newcommand{\RevJETP}{Sov. Phys. JETP}
\newcommand{\RevJETPrussian}{Zh. Eksp. Teor. Fiz.}
\newcommand{\Jprep}{Phys. Rep.}

\newcommand{\Jwrm}{Waves Rand. Media}

\newcommand{\VR}{V_R}

\newcommand{\tauS}{\tau_\mathrm{s}}
\newcommand{\tauB}{\tau^\star}

\newcommand{\lS}{l_\mathrm{s}}

\newcommand{\nbar}{\bar{n}}

\newcommand{\tProp}{t}

\newcommand{\deltavres}{\Delta v_\mathrm{res}}

\newcommand{\Dpres}{\Delta p_\mathrm{res}}

\begin{document}

\title{Coherent Backscattering of Ultracold Atoms}

\author{F. Jendrzejewski}
\author{K. M\"uller}
\author{J. Richard}
\author{A. Date}
\author{T. Plisson}
\affiliation{Laboratoire Charles Fabry UMR 8501,
Institut d'Optique, CNRS, Univ Paris Sud 11,
2 Avenue Augustin Fresnel,
91127 Palaiseau cedex, France}
\author{P. Bouyer}
\affiliation{LP2N UMR 5298,
Univ Bordeaux 1, Institut d'Optique and CNRS,
351 cours de la Lib\'eration,
33405 Talence, France.}
\author{A. Aspect}
\author{V. Josse}
\email{vincent.josse@institutoptique.fr}
\affiliation{Laboratoire Charles Fabry UMR 8501,
Institut d'Optique, CNRS, Univ Paris Sud 11,
2 Avenue Augustin Fresnel,
91127 Palaiseau cedex, France}

\date{\today}

\begin{abstract}
We report on the direct observation of coherent backscattering (CBS) of ultracold atoms in a quasi-two-dimensional configuration. Launching atoms with a well-defined momentum in a laser speckle disordered potential, we follow the progressive build up of the momentum scattering pattern, consisting of a ring associated with multiple elastic scattering, and the CBS peak in the backward direction. Monitoring the depletion of the initial momentum component and the formation of the angular ring profile allows us to determine microscopic transport quantities. We also study the time evolution of the CBS peak and find it in  fair agreement with predictions, at long times as well as at short times. The observation of CBS can be considered a direct signature of coherence in quantum transport of particles in disordered media. It is responsible for the so called weak localization phenomenon, which is the precursor of Anderson localization.
\end{abstract}


\maketitle

\begin{figure}[b]
\begin{center}
\includegraphics[width=0.96\linewidth]{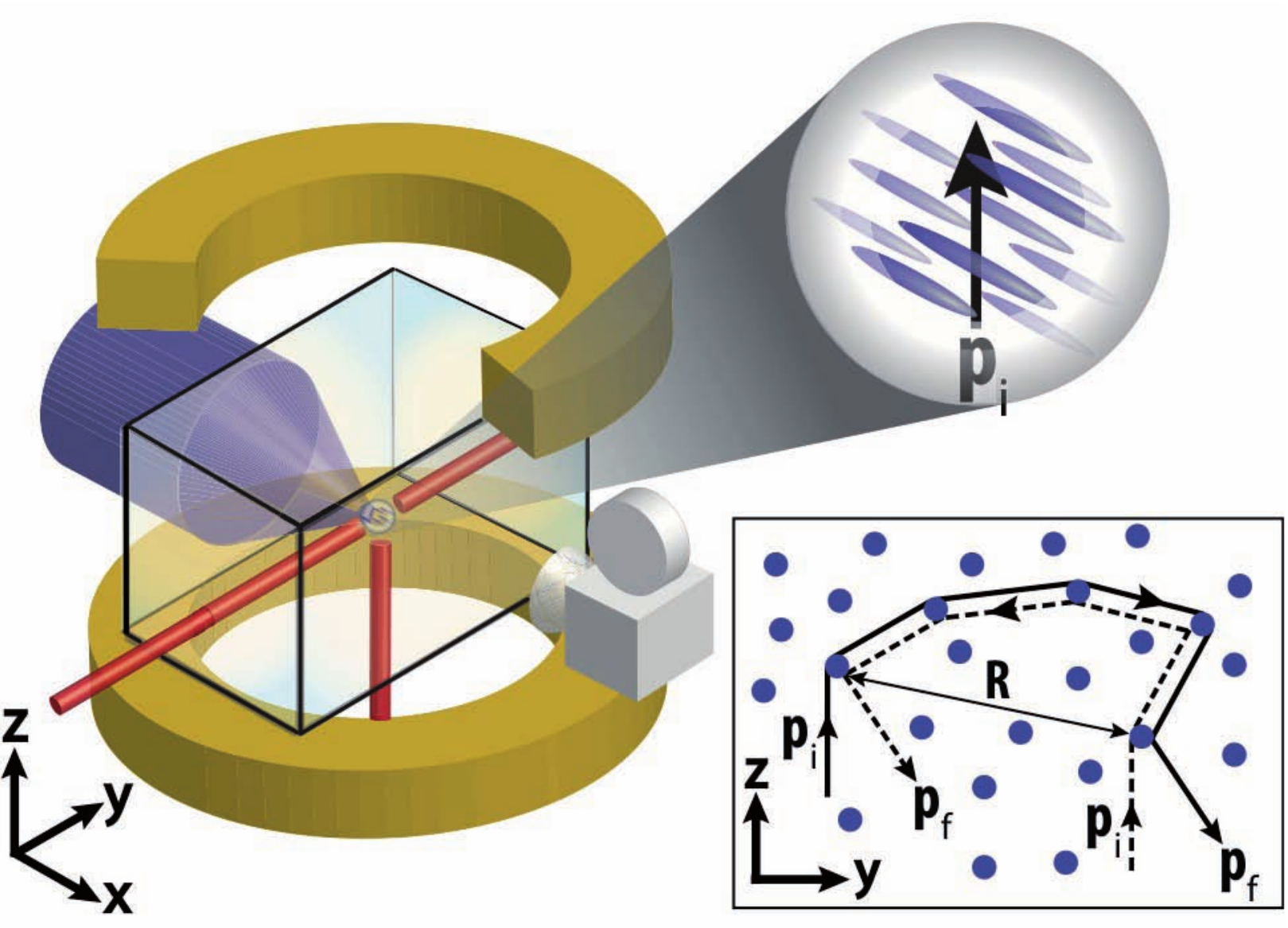}
\caption{Experimental set-up. A cloud of noninteracting ultracold atoms, released from an optical trap (beams along $y$ and $z$ axis, red) and suspended against gravity by magnetic levitation (horizontal coils, yellow), is launched with a well-defined momentum $\mathbf{p_\mathrm{i}}$ along the $z$ axis. It is submitted to an anisotropic laser speckle disordered potential (blue beam), propagating along the $x$ axis and elongated along that direction, leading to a quasi-2D diffusive motion in the $y-z$ plane (see text). The atomic momentum distribution in this plane is monitored by fluorescence imaging after a time of flight of~150 ms. Inset: physical origin of CBS. The coherent enhancement of scattering in the backward direction originates from the interference between each multiple scattering path (solid line) and its reversed counterpart (dashed line).}
\label{fig:Experiment}
\end{center}
\end{figure}

Quantum transport differs from classical transport by the crucial role of coherence effects. In the case of transport in a disordered medium, it can lead to the complete cancelling of transport when the disorder is strong enough: this is the celebrated Anderson localization (AL)~\cite{Anderson1958}. For weak disorder, a first order manifestation of coherence is the phenomenon of coherent backscattering (CBS), i.e., the enhancement of the scattering probability in the backward direction, due to a quantum interference of amplitudes associated with two opposite multiple scattering paths~\cite{Watson1969,Tsang1984,Akkermans1986} (see inset of Fig.~\ref{fig:Experiment}). Direct observation of such a peak is a smoking gun of the role of quantum coherence in quantum transport in disordered media.

CBS has been observed with classical waves in optics~\cite{Kuga1984,Albada1985,Wolf1985,Labeyrie1999}, acoustics~\cite{Bayer1993,Tourin1997}, and even seismology \cite{Larose2004}. In condensed matter physics, CBS is the basis of the weak localization phenomenon (see e.g.~\cite{ Altshuler1982}), which is responsible for the anomalous resistance of thin metallic films and its variation with an applied magnetic field~\cite{Altshuler1980,Bergmann1984}. In recent years, it has been possible to directly observe Anderson localization with ultracold atoms in one dimension~\cite{Billy2008,Roati2008a} and three dimensions~\cite{Kondov2011,Jendrzejewski2012a}. Convincing as they are, none of these experiments includes a direct evidence of the role of coherence.

\begin{figure}
\begin{center}
\includegraphics[width=1\linewidth]{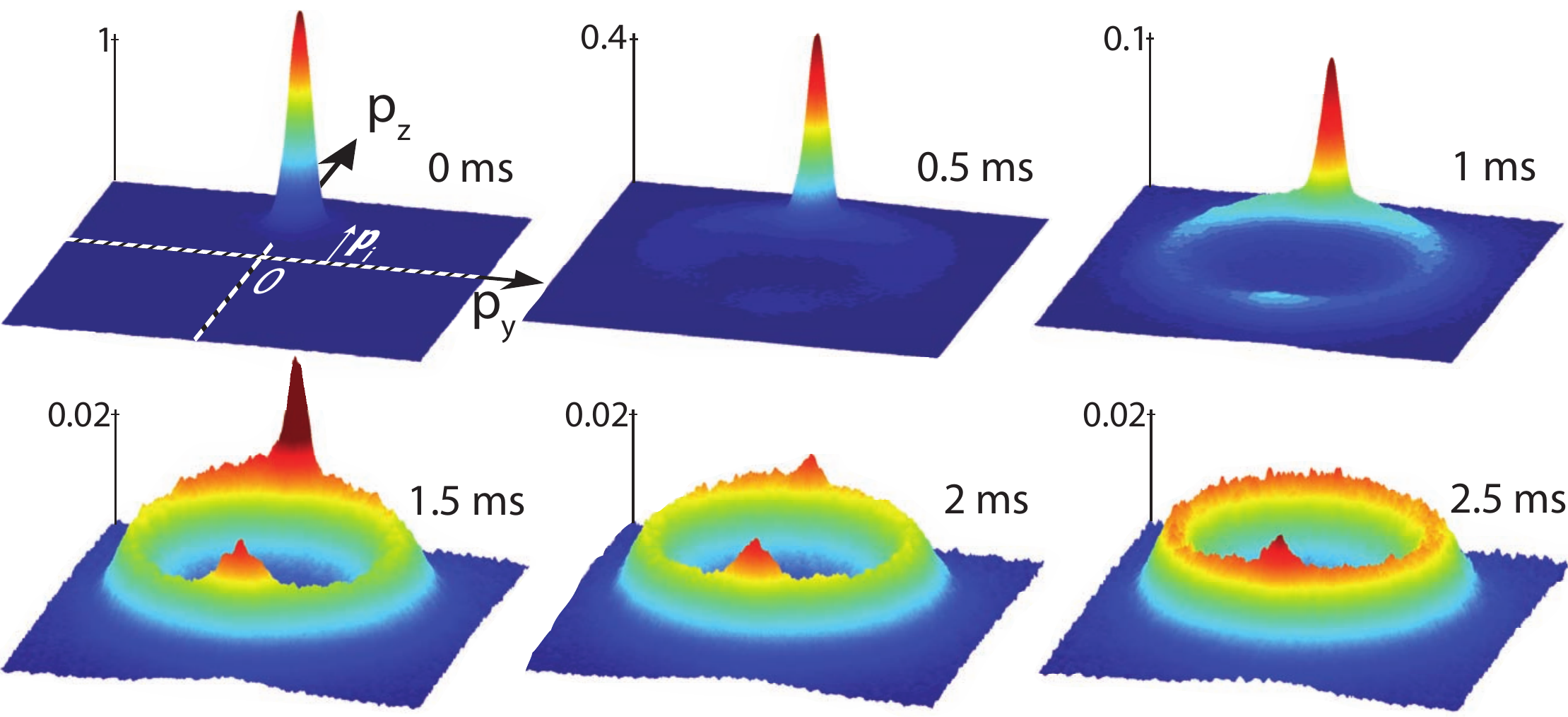}
\caption{Observed momentum distributions after different propagation times $t$ in the disorder. The images correspond to an averaging over 20 experimental runs.  Note that the vertical scale is different in the three first images  ($t=$0, 0.5, and 1~ms), whereas it is the same in the three last images  ($t=$1.5, 2, and 2.5~ms).}
 \label{fig:MomDist}
\end{center}
\end{figure}

In this Letter, we report on the direct observation of CBS with ultracold atoms, in a quasi-two-dimensional (2D) configuration~\cite{NoteLab}.
Our scheme is based on the proposal of Ref.~\cite{Cherroret2012} that suggested observing CBS in the momentum space.  A cloud of noninteracting ultracold atoms is launched with a narrow velocity distribution in a laser speckle disordered potential (Fig.~\ref{fig:Experiment}). Time of flight imaging, after propagation time $t$ in the disorder, directly yields the momentum distribution, as shown in Fig.~\ref{fig:MomDist}. As expected for elastic scattering of particles, we observe a ring that corresponds to a redistribution of the momentum directions over $2\pi$ while the momentum magnitude remains almost constant. The evolution of the initial momentum peak and of the angular ring profile yields the elastic scattering time and the transport time.
 But the most remarkable feature is the large visibility peak, which builds up in the backward direction. The height and width of that peak, and their evolution with time, are an indisputable signature of CBS, intimately linked to the role of coherence.

To understand the origin of that CBS peak, let us consider an input plane matter wave with initial momentum $\mathbf{p}_\mathrm{i}$ that experiences multiple scattering towards a final momentum $\mathbf{p}_\mathrm{f}$ (inset of Fig.~\ref{fig:Experiment}). For each multiple scattering path, we can consider the reversed path with the same input $\mathbf{p}_\mathrm{i}$ and output $\mathbf{p}_\mathrm{f}$. Since the initial and final atomic states are the same,  we must add the two corresponding complex quantum amplitudes, whose  phase difference is $\delta \phi= (\mathbf{p}_\mathrm{i}+\mathbf{p}_\mathrm{f})\cdot  \mathbf{R}/\hbar$ ($\mathbf{R}$ is the spatial separation between the initial and final scattering events and $\hbar=h/2\pi$ the reduced Planck constant). For the \emph{exact} backward momentum $\mathbf{p}_\mathrm{f}=-\mathbf{p}_\mathrm{i}$, the interference is always perfectly constructive, whatever the considered multiple scattering path. This coherent effect survives  the ensemble averaging over the disorder, so that the total scattering probability is twice as large as it would be in the incoherent case. For an increasing difference between $\mathbf{p}_\mathrm{f}$ and $-\mathbf{p}_\mathrm{i}$, the interference pattern is progressively washed out as we sum over all interference patterns associated with all multiple scattering paths. It results in a CBS peak of width inversely proportional to the spread $\Delta \mathbf{R}$ in the separations~\cite{NoteSeparation}. For diffusive scattering paths, the distribution of $\mathbf{R}$ is a Gaussian whose widths increase with time as $t^{1/2}$, and the CBS widths decrease  according to $\Delta p_{\mathrm{CBS},\alpha}=\hbar /\sqrt{2D_\alpha t}$ for each direction of space ($\alpha={y,z}$), $D_\alpha$ being the diffusion constant along that direction. This time resolved dynamics of the CBS peak has been observed in acoustics~\cite{Bayer1993,Tourin1997} and optics~\cite{Vreeker1988,NoteCusp}.

The crux of the experiment is a sample of noninteracting paramagnetic atoms, suspended against gravity by a magnetic gradient (as in~\cite{Jendrzejewski2012a}), and launched along the $z$ axis  with a very well-defined initial momentum $\mathbf{p}_\mathrm{i}$  (see Fig.~\ref{fig:Experiment}). This is realized in four steps. First,  evaporative cooling of an atomic cloud of $^{87}\mathrm{Rb}$ atoms in a quasi-isotropic optical dipole trap (trapping frequency~$\simeq5$~Hz) yields a  Bose-Einstein condensate of $9\times 10^4$ atoms in the $F = 2$, $m_F = -2$ ground sublevel. Second, we suppress the interatomic interactions by releasing the atomic cloud and letting it expand during 50 ms. At this stage, the atomic cloud has a size~(standard half-width along each direction) of  $\Delta r_\alpha =30$~$\mu$m, and the residual interaction energy ($E_\mathrm{int}/h\sim 1$~Hz) is negligible compared to all relevant energies of the problem. Since the atomic cloud is expanding radially with velocities proportional to the distance from the origin,  we can  use the  ``delta-kick cooling" technique~\cite{Ammann1997}, by switching on a harmonic potential for a well chosen amount of time. This almost freezes the motion of the atoms, and the resulting velocity spread  $\Delta v_\alpha= 0.12 \pm 0.03$~mm/s  is just one magnitude above the Heisenberg limit ($\Delta r_\alpha m \Delta v_\alpha  \sim 5\hbar$, with $m$ the atom mass). Last, we give the atoms a finite momentum $\mathbf{p}_\mathrm{i}$ along the $z$ direction, without changing the momentum spread, by applying an additional magnetic gradient during 12~ms. The first image of Fig.~\ref{fig:MomDist} shows the resulting 2D momentum distribution. The average velocity is  $v_\mathrm{i} = 3.3\pm 0.2$~mm/s ($k_\mathrm{i} =p_\mathrm{i}/\hbar\simeq 4.5$~$\mu$m$^{-1}$), corresponding to a kinetic energy $E_\mathrm{K} = p_\mathrm{i}^2/2m$ ($E_\mathrm{K}/h\simeq 1190$~Hz). This momentum distribution is obtained with a standard time of flight technique  that converts the  velocity distribution into a position distribution. Because of the magnetic levitation, we can let the atomic cloud expand ballistically for as long as 150 ms before performing fluorescence imaging along the $x$ axis. The overall velocity resolution of our experiment that takes into account the initial momentum spread, which writes $\deltavres=[{\Delta v_\alpha}^2+(\Delta r_\alpha /t_\mathrm{tof})^2]^{1/2}=0.23$ mm/s, is nevertheless mainly limited by the size $\Delta r_\alpha$ of the atomic cloud.

To study CBS, we suddenly switch on an optical disordered potential in less than 0.1~ms, let the atoms scatter for a time~$\tProp$, then switch off the disorder and monitor the momentum distribution at time $t$.
The disordered potential is the dipole potential associated with a laser speckle field \cite{goodman2007speckle,Clement2006}, obtained by passing a laser beam through a rough plate, and  focusing it on the atoms (Fig.~\ref{fig:Experiment}). It has an average value $V_\mathrm{R}$ (the disorder "amplitude") equal to its standard deviation. Its autocorrelation function is anisotropic, with a  transverse shape well represented by a Gaussian of standard half-widths $\sigma_y=\sigma_z=\sigma_\perp \simeq 0.2~\mu$m, and a longitudinal Lorentzian profile of half-width $\sigma_x \simeq 1~\mu$m (HWHM) \cite{Piraud2011a}. The laser (wavelength 532~nm) is detuned far off-resonance (wavelength 780~nm), yielding a purely conservative and repulsive potential. The disorder  amplitude $V_\mathrm{R}$ is homogenous to better than 1$\%$ over the atom cloud (profile of half-widths 1.2~mm along $y$,$z$, 1~mm along $x$).

The anisotropy of the speckle autocorrelation function (elongated along $x$) allows us to operate in a quasi-two-dimensional configuration by launching the atoms perpendicularly to the $x$ axis (along the $z$ axis).  In the  $y$-$z$ plane, the atoms are scattered by a potential with a correlation length  shorter than the reduced atomic de Broglie wavelength ($k_\mathrm{i} \sigma_\perp\simeq 0.9 $), so that the scattering probability is quasi-isotropic, and we will replace the subscript $\alpha = y,z$ by $\perp$ in the rest of this Letter. The dynamics within this plane develops on the typical time scale of a single scattering event, that is, the elastic scattering time $\tauS$. In contrast, the correlation length along the $x$ axis is  larger than the reduced atomic de Broglie wavelength ($k_\mathrm{i} \sigma_x \simeq 4.5$), so that each scattering event produces but a small deviation out of the $y$-$z$ plane~\cite{Kuhn2007a}. The diffusive dynamics along $x$ is then slower than in the $y-z$ plane, and for short times the dynamics is quasi 2D.

\begin{figure}[b]
\begin{center}
\includegraphics[width=1\linewidth]{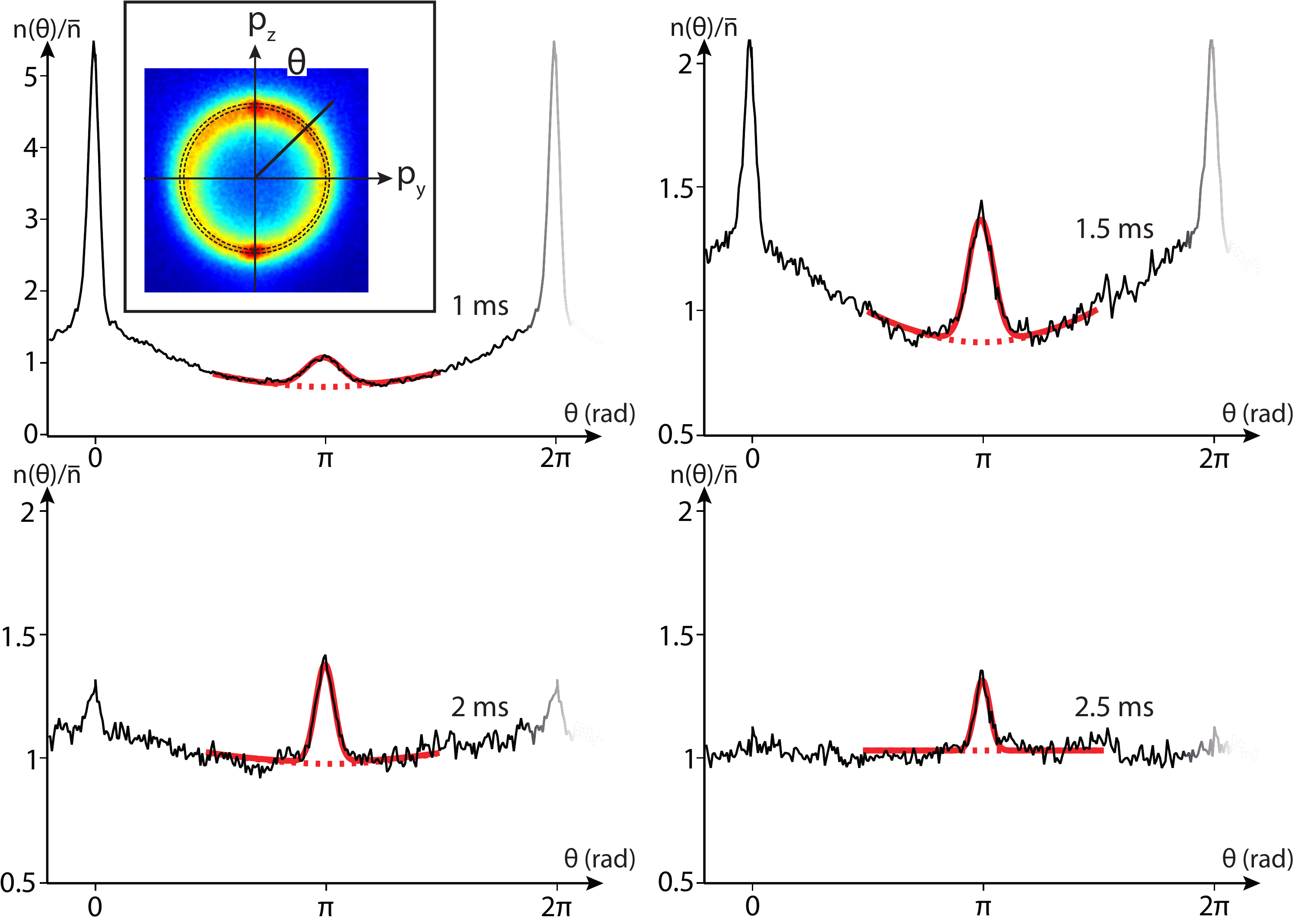}
\caption{Normalized angular profiles $n(\theta,t)/\bar{n}$ corresponding to the distributions shown in Fig.~\ref{fig:MomDist}. The red solid lines correspond to the double structure fit around the backscattering direction [a parabola for the incoherent background (dashed line) and a Gaussian for the CBS peak, see text]. Inset: false color representation of the momentum distribution ($t=2$~ms). The angle $\theta$ refers to the scattering direction with respect to the initial direction  ($\mathbf{p}_\mathrm{i}$).  }
\label{fig:Angular}
\end{center}
\end{figure}

Figure~\ref{fig:MomDist} shows  the time evolution of the momentum distribution for a disorder amplitude $V_\mathrm{R}/h = 975\pm 80$~Hz. In order to analyze these data quantitatively, we perform a radial integration of the 2D momentum distribution on a thin stripe between $p_\mathrm{i}-\Dpres$ and $p_\mathrm{i}+\Dpres$ (inset of Fig.~\ref{fig:Angular}) ($\Dpres=m \Delta v_\mathrm{res}$ is the momentum resolution). This yields the angular profile $n(\theta,t)$, displayed in Fig.~\ref{fig:Angular} for increasing diffusion times.

We first extract the elastic scattering time $\tauS$ from the exponential decay of the initial peak $n(\theta = 0,t) \propto e^{-t/\tauS}$~\cite{Akkermans2007book}. We find $\tauS = 0.33 \pm 0.02$~ms (mean free path $l_\mathrm{s}=v_\mathrm{i}\tauS=1.1$~$\mu$m). Note that this quantity also plays a role in the radial width of the ring (inset of Fig.~\ref{fig:Angular}), which is associated to a Lorentzian energy spread $\Delta E_\mathrm{dis}=\hbar/\tauS$ (HWHM) acquired by the atoms when the disorder is suddenly switched on. Combining the corresponding momentum spread with the resolution of our measurement, we find a width in agreement with the observed ring width. The measured value of $\tauS$ is in accordance with numerical simulations adapted to our configuration, but is about 2 times the value predicted by a perturbative calculation~\cite{Piraud2011a}. This is consistent with the fact that we are not fully in the weak disorder regime defined by $\Delta E_\mathrm{dis}/ E_\mathrm{K}=2/k_\mathrm{i}\lS\ll 1$. Here we have $\Delta E_\mathrm{dis}/E_\mathrm{K}\sim0.4$ ($k_\mathrm{i}\lS\sim 5$)~\cite{NoteSpectral}.

Monitoring the isotropization of the momentum distribution, we obtain another important quantity: the transport time $\tauB$, after which, in the absence of coherence, the information about the initial direction would be lost. It is determined from the exponential decay $e^{-t/\tauB}$ of the first component of the Fourier series expansion of $n(\theta,t)$~\cite{Cherroret2012, PrivateCord}. We find $\tauB = 0.4\pm 0.05$~ms, also in agreement with numerics. Note that $\tauB$ takes into account the CBS phenomenon and its calculation must include weak localization corrections. It is only slightly larger than $\tauS$, as expected for a nearly isotropic scattering probability ($k_\mathrm{i}\sigma_\perp=0.9$). The transport time sets the time scale of the onset of the diffusive dynamics, which is well established only after several  $\tauB$. In the experiment, we observe that the momentum distribution has become fully isotropic after $t\sim2.5$~ms (i.e. $\sim 6\, \tauB$), with a steady and flat angular profile of mean value $n(\theta,t)=\nbar$, except around $\theta = \pi$ where the CBS peak is still present.

\begin{figure}
\begin{center}
\includegraphics[width=1\linewidth]{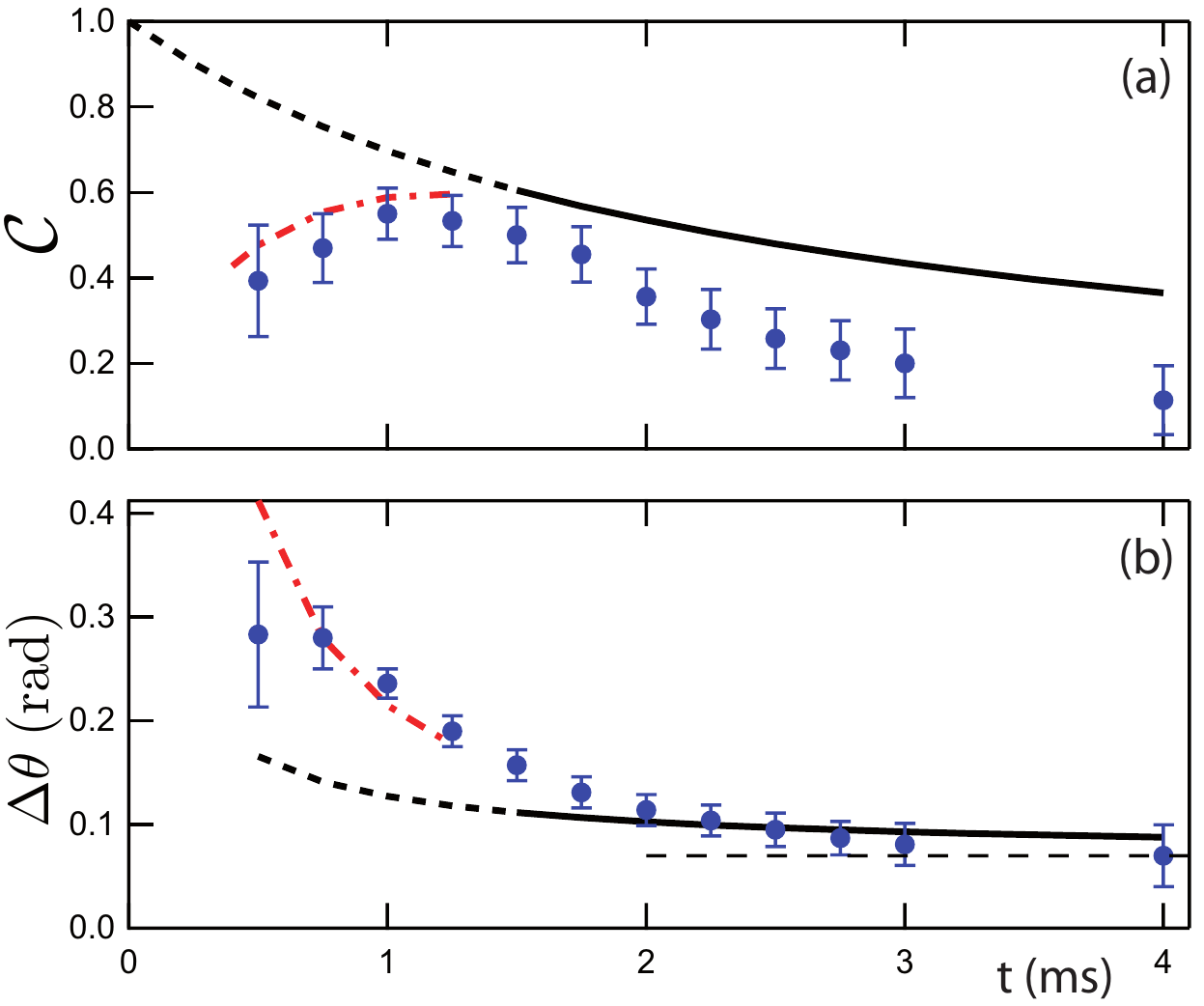}
\caption{Time resolved dynamics of the CBS peak: evolution of the contrast $\mathcal{C}$ (a) and the width  $\Delta \theta$ (b) versus the scattering time $t$. The blue points are experimental data and the error bars correspond to the 95$\%$ confidence intervals for the fitted values ($\pm$2 standard deviations). The theoretical predictions for the multiple scattering regime are represented by the solid black lines at long times ($t\gtrsim 4 \tauB$; $\tauB=0.4$~ms), and by dotted black lines at short times. The dashed-dotted red lines correspond to the calculation of~\cite{Gorodnichev1994} at short times (i.e. for $t\sim\tauS$; $\tauS=0.33$~ms), where single scattering events cannot be neglected.}
\label{fig:CBSevolv}
\end{center}
\end{figure}

To analyze the evolution of the CBS signal, we fit (see Fig.~\ref{fig:Angular}) the angular profiles by the function $n_{\mathrm{incoh}}(\theta,t)+ n_\mathrm{coh}(t)\exp[-(\theta-\pi)^2/2\Delta \theta(t)^2]$. In this formula, $n_{\mathrm{incoh}}$ is (within higher order terms) the multiple scattering background that would be obtained in the absence of coherence~\cite{Akkermans2007book2} and we assume that $n_{\mathrm{incoh}}(\theta,t)$ has a parabola shape around $\theta=\pi$. The fit allows us to measure the contrast $\mathcal{C}(t)=n_\mathrm{coh}(t)/n_{\mathrm{incoh}}(\pi,t)$ and the width $\Delta \theta (t)$ of the CBS peak, and to compare them to the results of theoretical predictions. Their evolutions are shown in Fig.~\ref{fig:CBSevolv}. A CBS peak appears as soon as scattering in the backward direction is significant, but the contrast starts decreasing before reaching the ideal value of 1. For the width, we observe the predicted monotonic decrease, but it tends asymptotically towards a non-null value rather than zero.

To compare these observations with theoretical predictions, we must  take into account our finite resolution. The black line in Fig.~\ref{fig:CBSevolv}(b) represents the calculated CBS width that results from the convolution of our resolution $\Delta \theta_\mathrm{res}=\Delta p_\mathrm{res}/p_\mathrm{i}=0.07$ with the expected CBS width $\Delta \theta_\mathrm{CBS}=\Delta p_{\mathrm{CBS},\perp}/p_\mathrm{i}=\hbar/p_\mathrm{i}\sqrt{2D_\perp t}$ in the diffusive regime (the widths are added quadratically). The diffusion constant is evaluated using the standard relation $D_\perp=v_\mathrm{i}^2 \tauB/2$, so that the solid line does not involve any adjustable parameter. We see that the agreement with the data is good when we enter the multiple scattering regime [for $t\gtrsim 1.5$~ ms in Fig.~\ref{fig:CBSevolv}(b)], but not at short times.

The broadening of the CBS peak by the finite resolution is also responsible for a decrease of the contrast, as represented in Fig.~\ref{fig:CBSevolv}(a). Here also, we observe that the prediction (which again involves no adjustable parameter) is very different from the observed values at short times, but is in reasonable agreement with the measurements around $t\sim1.5$~ms. On the other hand, the measured contrast is definitely smaller than the theoretical prediction when $t$ increases yet more.
We relate this observation at long times  to the onset of the dynamics in the third direction $x$. Using a second imaging system yielding the momentum distribution in the $x-y$ plane, we estimate a typical time of 4 ms for this out-of-plane dynamics to become significant, and render the 2D approach wanting.

Deviations at short times were to be expected. Indeed, CBS demands multiple scattering, or at least double scattering, to happen (see inset of Fig.~\ref{fig:Experiment}), whereas single scattering events do not participate to the CBS peak. At short times ($t\sim\tauS$), the contribution of single scattering to backscattering is not negligible compared to multiple scattering. This entails a reduction of the contrast (see e.g.~\cite{Tsang1984}), and a modification of the shape (no longer Gaussian), whose width decreases at this stage as $1/t$ (ballistic motion between the first two scatterers). In the case of light, a calculation for isotropic scattering~\cite{Gorodnichev1994} predicts a short time evolution of the contrast $\mathcal{C}=(2t/\pi\tauS)/(1+2t/\pi\tauS)$ and width $\Delta \theta_\mathrm{CBS}\sim 3/k_\mathrm{i} \lS (\tauS/t)$. This prediction is plotted in Fig.~\ref{fig:CBSevolv} and is found in fair agreement with the observations in this time domain. Finally, note that the width around $t\sim\tauS$ is linked to the disorder strength quantified by $k_\mathrm{i}\lS$. Here we find a maximum value of $\Delta \theta_\mathrm{max}\sim0.3$ rad, that is $\Delta \theta_\mathrm{max}\sim 1.5/k_\mathrm{i}\lS$ ($k_\mathrm{i}\lS\sim 5$, see above).

Similar measurements and analysis have been repeated for weaker disorder ($\VR/h=525,750\mathrm{~Hz}$) and smaller initial momentum $p_\mathrm{i}$ ($E_\mathrm{K}/h=160, 220, 620\mathrm{~Hz}$), and we have found a similar agreement between data and theory. In contrast to the observed moderate changes in the maximum peak contrast ($\mathcal{C_\mathrm{max}}\sim0.5-0.7$), the  maximum peak width $\Delta \theta_\mathrm{max}$ increases significantly with the amplitude of the disorder and the inverse of $p_\mathrm{i}$. The highest observed width of $1.2$~rad (from which we infer $k_\mathrm{i}\lS\sim1.25$) suggests that we are very close to the strong disorder regime, where AL is expected to be experimentally observable in 2D systems. Such an observation, however, would demand a longer 2D evolution in the disorder, which is limited in the present experiment because of the cross over to the 3D regime. Increasing this time, as for instance in~\cite{Robert2010}, will then constitute the next step towards AL, with the possibility to observe the coherent forward scattering peak predicted in~\cite{Karpiuk2012}.

In conclusion, we have demonstrated experimentally that the time resolved study of the momentum distribution of ultracold atoms in a random potential is a powerful tool to study quantum transport properties in disordered media. We have been able to extract the elastic scattering time, the transport time, and to observe and study the evolution of the CBS peak. Let us emphasize that the theoretical analysis as well as numerical simulations render an account of the observations not only in the multiple scattering regime but also at short time, during the onset of multiple scattering. Such agreement gives a strong evidence of the fundamental role of coherence in that phenomenon. Further evidences of the role of coherence could be sought in the predicted suppression of the CBS peak~\cite{Golubentsev1984} when scrambling the disorder, or when dephasing the counter-propagating multiple scattering paths using  artificial gauge fields~\cite{Lin2009b}, in the spirit of  pioneering works in condensed matter physics~\cite{Bergmann1984} or optics~\cite{Lenke2000}. Finally, this work also opens promising prospects to study the effect of interactions on  CBS (see e.g.~\cite{Agranovich1991,Hartung2008}).

\begin{acknowledgments}
We thank T. Bourdel, C. M\"uller and B. van Tiggelen for fruitful discussions and comments. This research was supported by
 ERC (Advanced Grant "Quantatop"),
 ANR (ANR-08-blan-0016-01),
 MESR,
DGA,
RTRA Triangle de la Physique, IXBLUE and IFRAF.
\end{acknowledgments}


\begin{thebibliography}{0}%
\makeatletter
\providecommand \@ifxundefined [1]{%
 \@ifx{#1\undefined}
}%
\providecommand \@ifnum [1]{%
 \ifnum #1\expandafter \@firstoftwo
 \else \expandafter \@secondoftwo
 \fi
}%
\providecommand \@ifx [1]{%
 \ifx #1\expandafter \@firstoftwo
 \else \expandafter \@secondoftwo
 \fi
}%
\providecommand \natexlab [1]{#1}%
\providecommand \enquote  [1]{``#1''}%
\providecommand \bibnamefont  [1]{#1}%
\providecommand \bibfnamefont [1]{#1}%
\providecommand \citenamefont [1]{#1}%
\providecommand \href@noop [0]{\@secondoftwo}%
\providecommand \href [0]{\begingroup \@sanitize@url \@href}%
\providecommand \@href[1]{\@@startlink{#1}\@@href}%
\providecommand \@@href[1]{\endgroup#1\@@endlink}%
\providecommand \@sanitize@url [0]{\catcode `\\12\catcode `\$12\catcode
  `\&12\catcode `\#12\catcode `\^12\catcode `\_12\catcode `\%12\relax}%
\providecommand \@@startlink[1]{}%
\providecommand \@@endlink[0]{}%
\providecommand \url  [0]{\begingroup\@sanitize@url \@url }%
\providecommand \@url [1]{\endgroup\@href {#1}{\urlprefix }}%
\providecommand \urlprefix  [0]{URL }%
\providecommand \Eprint [0]{\href }%
\providecommand \doibase [0]{http://dx.doi.org/}%
\providecommand \selectlanguage [0]{\@gobble}%
\providecommand \bibinfo  [0]{\@secondoftwo}%
\providecommand \bibfield  [0]{\@secondoftwo}%
\providecommand \translation [1]{[#1]}%
\providecommand \BibitemOpen [0]{}%
\providecommand \bibitemStop [0]{}%
\providecommand \bibitemNoStop [0]{.\EOS\space}%
\providecommand \EOS [0]{\spacefactor3000\relax}%
\providecommand \BibitemShut  [1]{\csname bibitem#1\endcsname}%
\let\auto@bib@innerbib\@empty
\end{thebibliography}%


\begin{thebibliography}{10}
\expandafter\ifx\csname url\endcsname\relax
  \def\url#1{\texttt{#1}}\fi
\expandafter\ifx\csname urlprefix\endcsname\relax\def\urlprefix{URL }\fi
\providecommand{\bibinfo}[2]{#2}
\providecommand{\eprint}[2][]{\url{#2}}

\bibitem{Anderson1958}
\bibinfo{author}{P. W. Anderson},
\newblock \bibinfo{journal}{\Jpr} \textbf{\bibinfo{volume}{109}},
  \bibinfo{pages}{1492} (\bibinfo{year}{1958}).

\bibitem{Watson1969}
\bibinfo{author}{K. M. Watson},
\newblock \bibinfo{journal}{\Jmp} \textbf{\bibinfo{volume}{10}},
  \bibinfo{pages}{688} (\bibinfo{year}{1969});
\bibinfo{author}{D. A. de Wolf},
\newblock \bibinfo{journal}{\JTAP} \textbf{\bibinfo{volume}{19}},
  \bibinfo{pages}{254} (\bibinfo{year}{1971});
\bibinfo{author}{Yu. N. Barabanenkov},
\newblock \bibinfo{journal}{\JRQE} \textbf{\bibinfo{volume}{16}},
  \bibinfo{pages}{65} (\bibinfo{year}{1973}).

\bibitem{Tsang1984}
\bibinfo{author}{L. Tsang} and \bibinfo{author}{A. Ishimaru},
\newblock \bibinfo{journal}{\JOSAA} \textbf{\bibinfo{volume}{1}},
\bibinfo{pages}{836} (\bibinfo{year}{1984}).

\bibitem{Akkermans1986}
\bibinfo{author}{E. Akkermans}, \bibinfo{author}{P. E. Wolf}, and \bibinfo{author}{R. Maynard},
\newblock \bibinfo{journal}{\Jprl} \textbf{\bibinfo{volume}{56}},
  \bibinfo{pages}{1471} (\bibinfo{year}{1986}).

\bibitem{Kuga1984}
\bibinfo{author}{Y. Kuga} and \bibinfo{author}{A. Ishimaru},
\newblock \bibinfo{journal}{\JOSAA} \textbf{\bibinfo{volume}{1}},
  \bibinfo{pages}{831} (\bibinfo{year}{1984}).

\bibitem{Albada1985}
\bibinfo{author}{M. P. Van Albada} and \bibinfo{author}{A. Lagendijk},
\newblock \bibinfo{journal}{\Jprl} \textbf{\bibinfo{volume}{55}},
  \bibinfo{pages}{2692} (\bibinfo{year}{1985}).

\bibitem{Wolf1985}
\bibinfo{author}{P. E. Wolf} and \bibinfo{author}{G. Maret},
\newblock \bibinfo{journal}{\Jprl} \textbf{\bibinfo{volume}{55}},
\bibinfo{pages}{2696} (\bibinfo{year}{1985}).

\bibitem{Labeyrie1999}
\bibinfo{author}{G. Labeyrie}, \bibinfo{author}{F. de Tomasi}, \bibinfo{author}{J. C. Bernard}, \bibinfo{author}{C. A. M\"{u}ller}, \bibinfo{author}{C. Miniatura}, and \bibinfo{author}{R. Kaiser},
\newblock \bibinfo{journal}{\Jprl} \textbf{\bibinfo{volume}{83}},
\bibinfo{pages}{5266} (\bibinfo{year}{1999}).

\bibitem{Bayer1993}
\bibinfo{author}{G. Bayer} and \bibinfo{author}{T. Niederdr\"{a}nk},
\newblock \bibinfo{journal}{\Jprl} \textbf{\bibinfo{volume}{70}},
\bibinfo{pages}{3884} (\bibinfo{year}{1993}).

\bibitem{Tourin1997}
\bibinfo{author}{A. Tourin}, \bibinfo{author}{A. Derode}, \bibinfo{author}{P. Roux}, \bibinfo{author}{B. A. van Tiggelen},  and \bibinfo{author}{M. Fink},
\newblock \bibinfo{journal}{\Jprl} \textbf{\bibinfo{volume}{79}},
\bibinfo{pages}{3637} (\bibinfo{year}{1997}).

\bibitem{Larose2004}
\bibinfo{author}{E. Larose}, \bibinfo{author}{L. Margerin}, \bibinfo{author}{B. A. van Tiggelen}, and \bibinfo{author}{M. Campillo},
\newblock \bibinfo{journal}{\Jprl} \textbf{\bibinfo{volume}{93}},
\bibinfo{pages}{048501} (\bibinfo{year}{2004}).

\bibitem{Altshuler1982}
\bibinfo{author}{B. L. Altshuler}, \bibinfo{author}{A. G. Aronov}, \bibinfo{author}{D. E. Khmelnitskii}, and \bibinfo{author}{A. I. Larkin},
\newblock \emph{\bibinfo{booktitle}{Quantum Theory of Solids}}
\newblock (\bibinfo{publisher}{Mir, Moscow} \bibinfo{year}{1982}), p. \bibinfo{pages}{130}.

\bibitem{Altshuler1980}
\bibinfo{author}{B. L. Altshuler}, \bibinfo{author}{D. Khmelnitzkii}, and \bibinfo{author}{A. I. Larkin}, and \bibinfo{author}{P. A. Lee},
\newblock \bibinfo{journal}{\Jprb} \textbf{\bibinfo{volume}{22}},
\bibinfo{pages}{5142} (\bibinfo{year}{1980}).


\bibitem{Bergmann1984}
\bibinfo{author}{G. Bergmann},
\newblock \bibinfo{journal}{\Jprep} \textbf{\bibinfo{volume}{107}},
\bibinfo{pages}{1} (\bibinfo{year}{1984}).

\bibitem{Billy2008}
\bibinfo{author}{J. Billy}, \bibinfo{author}{V. Josse}, \bibinfo{author}{Z. Zuo}, \bibinfo{author}{A. Bernard}, \bibinfo{author}{B. Hambrecht}, \bibinfo{author}{P. Lugan}, \bibinfo{author}{D. Cl\'ement}, \bibinfo{author}{L. Sanchez-Palencia}, \bibinfo{author}{P. Bouyer}, and \bibinfo{author}{A. Aspect},
\newblock \bibinfo{journal}{\Jnature} \textbf{\bibinfo{volume}{453}},
  \bibinfo{pages}{891} (\bibinfo{year}{2008}).

\bibitem{Roati2008a}
\bibinfo{author}{G. Roati}, \bibinfo{author}{C. D'Errico}, \bibinfo{author}{L. Fallani}, \bibinfo{author}{M. Fattori}, \bibinfo{author}{C. Fort}, \bibinfo{author}{M. Zaccanti}, \bibinfo{author}{G. Modugno}, \bibinfo{author}{M. Modugno}, and  \bibinfo{author}{M. Inguscio},
\newblock \bibinfo{journal}{\Jnature} \textbf{\bibinfo{volume}{453}},
  \bibinfo{pages}{895} (\bibinfo{year}{2008}).

\bibitem{Kondov2011}
\bibinfo{author}{S. S. Kondov}, \bibinfo{author}{W. R. McGehee}, \bibinfo{author}{J. J. Zirbel}, and \bibinfo{author}{B. DeMarco},
\newblock \bibinfo{journal}{Science} \textbf{\bibinfo{volume}{334}},
  \bibinfo{pages}{66} (\bibinfo{year}{2011}).

\bibitem{Jendrzejewski2012a}
\bibinfo{author}{F. Jendrzejewski}, \bibinfo{author}{A. Bernard}, \bibinfo{author}{K. M\"uller}, \bibinfo{author}{P. Cheinet}, \bibinfo{author}{V. Josse}, \bibinfo{author}{M. Piraud}, \bibinfo{author}{L. Pezz\'e}, \bibinfo{author}{L. Sanchez-Palencia}, \bibinfo{author}{A. Aspect}, and \bibinfo{author}{P. Bouyer},
\newblock \bibinfo{journal}{\Jnatphys} \textbf{\bibinfo{volume}{8}},
  \bibinfo{pages}{398} (\bibinfo{year}{2012}).

\bibitem{NoteLab}
\bibinfo{note}{During the preparation of this manuscript we have been made aware of an independent similar observation~\cite{Labeyrie2012}. The incoherent backscattering echo phenomenon reported in that paper, which may hamper the observation of CBS, does not play a role in our case, since the delta kick cooling method that we use suppresses the position-momentum correlation in the atomic sample. Moreover,  the large amplitude of our suddenly applied disorder  would wash out any residual correlation.}

 \bibitem{Labeyrie2012}
\bibinfo{author}{G. Labeyrie}, \bibinfo{author}{T. Karpiuk}, \bibinfo{author}{B. Gr\'emaud}, \bibinfo{author}{C. Minitatura}, and \bibinfo{author}{D. Delande},
  \bibinfo {note} {arXiv:1206.0845}.

\bibitem{Cherroret2012}
\bibinfo{author}{N. Cherroret}, \bibinfo{author}{T. Karpiuk}, \bibinfo{author}{C. A. M\"uller}, \bibinfo{author}{B. Gr\'emaud}, and \bibinfo{author}{C. Miniatura},
\newblock \bibinfo{journal}{\Jpra} \textbf{\bibinfo{volume}{85}},
  \bibinfo{pages}{011604(R)} (\bibinfo{year}{2012}).

\bibitem{NoteSeparation}
\bibinfo{note}{More precisely, the peak at time $t$ is the Fourier transform of the 2D distribution of the separations $\mathbf{R}$ at time t.}

\bibitem{Vreeker1988}
\bibinfo{author}{R. Vreeker}, \bibinfo{author}{M. P. van Albada}, \bibinfo{author}{R. Sprik},  and \bibinfo{author}{A. Lagendijk},
\newblock \bibinfo{journal}{Phys. Lett. A} \textbf{\bibinfo{volume}{132}},
 \bibinfo{pages}{51} (\bibinfo{year}{1988}).

\bibitem{NoteCusp}
\bibinfo{note}{In the stationary regime, the CBS peak results from the sum of all the contributions for all possible diffusion times and has the celebrated cusp shape predicted in~\cite{Akkermans1986}, and observed in optics for instance in~\cite{Wiersma1995,Labeyrie1999}.}

\bibitem{Wiersma1995}
\bibinfo{author}{D. S. Wiersma}, \bibinfo{author}{M. P. van Albada},  \bibinfo{author}{B. A. van Tiggelen}, and \bibinfo{author}{A. Lagendijk},
\newblock \bibinfo{journal}{\Jprl} \textbf{\bibinfo{volume}{74}},
 \bibinfo{pages}{4193} (\bibinfo{year}{1995}).

\bibitem{Ammann1997}
\bibinfo{author}{H. Ammann} and \bibinfo{author}{N. Christensen},
\newblock \bibinfo{journal}{\Jprl} \textbf{\bibinfo{volume}{78}},
  \bibinfo{pages}{2088} (\bibinfo{year}{1997}).

\bibitem{Clement2006}
\bibinfo{author}{D. Cl\'ement}, \bibinfo{author}{A. F. Varon}, \bibinfo{author}{J. A. Retter}, \bibinfo{author}{L. Sanchez-Palencia}, \bibinfo{author}{A. Aspect}, and \bibinfo{author}{P. Bouyer},
\newblock \bibinfo{journal}{\Jnjp} \textbf{\bibinfo{volume}{8}},
  \bibinfo{pages}{165} (\bibinfo{year}{2006}).

\bibitem{goodman2007speckle}
\bibinfo{author}{J. W. Goodman}
\newblock \emph{\bibinfo{title}{{Speckle Phenomena in Optics: Theory and
  Applications}}} (\bibinfo{publisher}{Roberts and Co, Englewood}
  \bibinfo{year}{2007}).


\bibitem{Piraud2011a}
\bibinfo{author}{M. Piraud}, \bibinfo{author}{L. Pezz\'e}, and \bibinfo{author}{L. Sanchez-Palencia},
\newblock \bibinfo{journal}{\EPL}  \textbf{\bibinfo{volume}{99}},
  \bibinfo{pages}{50 003} (\bibinfo{year}{2012}).
.

\bibitem{Kuhn2007a}
\bibinfo{author}{R. C. Kuhn}, \bibinfo{author}{O. Sigwarth},  \bibinfo{author}{C. Miniatura}, \bibinfo{author}{D. Delande},  and  \bibinfo{author}{C. A. M\"uller},
\newblock \bibinfo{journal}{\Jnjp} \textbf{\bibinfo{volume}{9}},
  \bibinfo{pages}{161} (\bibinfo{year}{2007}).


\bibitem{Akkermans2007book}
\bibinfo{author}{E. Akkermans and G. Montambaux}
\newblock \emph{\bibinfo{title}{{Mesoscopic Physics of Electrons and Photons}}} (\bibinfo{publisher}{Cambridge University Press, Cambridge, England},
  \bibinfo{year}{2007}) \bibinfo{pages}.

\bibitem{NoteSpectral}
\bibinfo{note}{The shape of the energy distribution (which is intimately linked to the so-called \emph{spectral function}) departs from a Lorentzian when approaching the strong disorder regime~\cite{Yedjour2010}.}

\bibitem{Yedjour2010}
\bibinfo{author}{A. Yedjour}, and \bibinfo{author}{B. A. van Tiggelen},
\newblock \bibinfo{journal}{\EPJD} \textbf{\bibinfo{volume}{59}},
  \bibinfo{pages}{249} (\bibinfo{year}{2010}).

\bibitem{PrivateCord}
\bibinfo{author}{T. Plisson},  \bibinfo{author}{T. Bourdel}, and \bibinfo{author}{C. A. M\"uller},
\bibinfo{note}{arXiv:1209.1477.}

\bibitem{Akkermans2007book2}
\bibinfo{author}{E. Akkermans and G. Montambaux}
\newblock \emph{\bibinfo{title}{{Mesoscopic Physics of Electrons and Photons}}} (\bibinfo{publisher}{Cambridge University Press, Cambridge, England},
  \bibinfo{year}{2007}) \bibinfo{pages}{Chap. 8, Sec. 8.3.1. }


\bibitem{Gorodnichev1994}
\bibinfo{author}{E. E. Gorodnichev} and  \bibinfo{author}{D. B. Rogozkin},
\newblock \bibinfo{journal}{\Jwrm} \textbf{\bibinfo{volume}{4}},
  \bibinfo{pages}{51} (\bibinfo{year}{1994}).

   \bibitem{Robert2010}
\bibinfo{author}{M. Robert-de-Saint-Vincent}, \bibinfo{author}{J.-P. Brantut}, \bibinfo{author}{B. Allard}, \bibinfo{author}{T. Plisson}, \bibinfo{author}{L. Pezz\'e}, \bibinfo{author}{L. Sanchez-Palencia}, \bibinfo{author}{A. Aspect}, \bibinfo{author}{T. Bourdel},  and \bibinfo{author}{P. Bouyer},
\newblock \bibinfo{journal}{\Jprl} \textbf{\bibinfo{volume}{104}},
  \bibinfo{pages}{220602} (\bibinfo{year}{2010}).


\bibitem{Karpiuk2012}
\bibinfo{author}{T. Karpiuk}, \bibinfo{author}{N. Cherroret}, \bibinfo{author}{K. L. Lee}, \bibinfo{author}{B. Gr\'emaud}, \bibinfo{author}{C.A. M\"uller}, and \bibinfo{author}{C. Miniatura},
  \bibinfo {note} {arXiv:1204.3451}, \bibinfo {note} {[Phys. Rev. Lett. (to be published)]}.


 \bibitem{Golubentsev1984}
\bibinfo{author}{A. A. Golubentsev},
\newblock \bibinfo{journal}{\RevJETPrussian} \textbf{\bibinfo{volume}{86}},
  \bibinfo{pages}{47} (\bibinfo{year}{1984}) [\bibinfo{journal}{\RevJETP}, \textbf{\bibinfo{volume}{59}},
  \bibinfo{pages}{26} (\bibinfo{year}{1984})]. 


\bibitem{Lin2009b}
\bibinfo{author}{Y.-J. Lin}, \bibinfo{author}{R. L. Compton},  \bibinfo{author}{K. Jim\'enez-Garc\'ia}, \bibinfo{author}{J. V. Porto},  and  \bibinfo{author}{I. B. Spielman},
\newblock \bibinfo{journal}{\Jnature} \textbf{\bibinfo{volume}{462}},
  \bibinfo{pages}{628} (\bibinfo{year}{2009}).

\bibitem{Lenke2000}
\bibinfo{author}{R. Lenke} and \bibinfo{author}{G. Maret},
\newblock \bibinfo{journal}{\Jepb} \textbf{\bibinfo{volume}{17}},
  \bibinfo{pages}{171} (\bibinfo{year}{2000}).

\bibitem{Agranovich1991}
\bibinfo{author}{V. M. Agranovich} and \bibinfo{author}{V. E. Kravtsov},
\newblock \bibinfo{journal}{\Jprb} \textbf{\bibinfo{volume}{43}},
  \bibinfo{pages}{13691} (\bibinfo{year}{1991}).

\bibitem{Hartung2008}
\bibinfo{author}{M. Hartung}, \bibinfo{author}{T. Wellens}, \bibinfo{author}{C. A. M\"uller}, \bibinfo{author}{K. Richter},  and \bibinfo{author}{P. Schlagheck},
\newblock \bibinfo{journal}{\Jprl} \textbf{\bibinfo{volume}{101}},
  \bibinfo{pages}{020603} (\bibinfo{year}{2008}).

\end{thebibliography}
\end{document}